# Neuroscience needs Network Science

Dániel L Barabási[1,2,✉], Ginestra Bianconi[3,4], Ed Bullmore[5], Mark Burgess[6], SueYeon Chung[7,8], Tina Eliassi-Rad[9,10,11], Dileep George[12], István A. Kovács[13,14], Hernán Makse[15], Christos Papadimitriou[16], Thomas E. Nichols[17,18], Olaf Sporns[19], Kim Stachenfeld[12,16], Zoltán Toroczkai[20], Emma K. Towlson[21], Anthony M Zador[22], Hongkui Zeng[23], Albert-László Barabási[9,24,25], Amy Bernard[26], György Buzsáki[27,28,✉]

[1] Biophysics Program, Harvard University, Cambridge, MA, USA
[2] Department of Molecular and Cellular Biology and Center for Brain Science, Harvard University, Cambridge, Massachusetts, USA
[3] School of Mathematical Sciences, Queen Mary University of London, London, E1 4NS, UK
[4] The Alan Turing Institute, The British Library, London, NW1 2DB, UK
[5] Department of Psychiatry and Wolfson Brain Imaging Centre, University of Cambridge, Cambridge, United Kingdom
[6] ChiTek-i AS, Oslo, Norway
[7] Center for Neural Science, New York University, New York, NY, USA.
[8] Center for Computational Neuroscience, Flatiron Institute, Simons Foundation, New York, NY, USA.
[9] Network Science Institute, Northeastern University, Boston, MA, USA
[10] Khoury College of Computer Sciences, Northeastern University, Boston, MA, USA
[12] Santa Fe Institute, Santa Fe, NM, USA
[12] DeepMind
[13] Department of Physics and Astronomy, Northwestern University, 633 Clark Street, Evanston, IL 60208, USA
[14] Northwestern Institute on Complex Systems, Chambers Hall, 600 Foster St, Northwestern University, Evanston, IL 60208
[15] Levich Institute and Physics Department, City College of New York, New York, NY 10031 US
[16] Columbia University, New York, NY, 10027, USA
[17] Big Data Institute, Li Ka Shing Centre for Health Information and Discovery, Nuffield Department of Population Health, University of Oxford, Oxford, OX3 7LF, UK
[18] Wellcome Centre for Integrative Neuroimaging, FMRIB, Nuffield Department of Clinical Neurosciences, University of Oxford, Oxford, OX3 9DU, UK
[19] Department of Psychological and Brain Sciences, Indiana University, Bloomington IN 47405
[20] Department of Physics, University of Notre Dame, 225 Nieuwland Science Hall, Notre Dame IN 46556, USA
[21] Department of Computer Science, Department of Physics and Astronomy, Hotchkiss Brain Institute, Children's Research Hospital, University of Calgary, Calgary, Alberta, Canada
[22] Cold Spring Harbor Laboratory, Cold Spring Harbor, NY, 11724
[23] Allen Institute for Brain Science, Seattle, WA, USA.
[24] Department of Medicine, Brigham and Women's Hospital and Harvard Medical School, Boston, MA, 02115, USA
[25] Department of Network and Data Science, Central European University, Budapest, H-1051, Hungary
[26] The Kavli Foundation, Los Angeles, CA, USA
[27] Neuroscience Institute and Department of Neurology, NYU Grossman School of Medicine, New York University, New York, NY, USA
[28] Center for Neural Science, New York University, New York, NY, USA
✉ Correspondence: danielbarabasi@gmail.com (D.L.B.), Gyorgy.Buzsaki@nyulangone.org (Gy.B.)

**Abstract**

The brain is a complex system comprising a myriad of interacting elements, posing significant challenges in understanding its structure, function, and dynamics. Network science has emerged as a powerful tool for studying such intricate systems, offering a framework for integrating multiscale data and complexity. Here, we discuss the application of network science in the study of the brain, addressing topics such as network models and metrics, the connectome, and the role of dynamics in neural networks. We explore the challenges and opportunities in integrating multiple data streams for understanding the neural transitions from development to healthy function to disease, and discuss the potential for collaboration between network science and neuroscience communities. We underscore the importance of fostering interdisciplinary opportunities through funding initiatives, workshops, and conferences, as well as supporting students and postdoctoral fellows with interests in both disciplines. By uniting the network science and neuroscience communities, we can develop novel network-based methods tailored to neural circuits, paving the way towards a deeper understanding of the brain and its functions.

**Introduction**

During the past two decades, network science has become a vital tool in the study of complex systems, offering a wide range of analytical and algorithmic techniques to explore the structure of a complex, interconnected system. Previous reductionist approaches, built upon decades of empirical research, have focused on the functioning of individual elements while neglecting how their interactions give rise to emergent aspects of organization. More recently, network approaches helped map out the interactions between molecules, cells, tissues, individuals, and organizations. It is becoming clear that we need network theory in neuroscience to understand how distributed patterns of interactions create function, and to account for the complexity of integrated systems.

The brain, with its billions of cells connected by synapses, is the ultimate example of a complex system that cannot be understood through the study of individual components alone.

In order to unveil the neural basis of complex behaviors and functions such as perception, movement, cognition, memory, and emotion, we must acknowledge and catalog the interactions between the neurons, allowing us to integrate multiple levels of observations and apply diverse approaches, including computational and mathematical modeling (Pulvermüller et al. 2021; Bassett, Zurn, and Gold 2018; Fornito, Zalesky, and Breakspear 2015). The goal of this paper is to outline how network science is not only well-suited, but is necessary to the integration of our knowledge of individual neurons into a broader understanding of brain function. Towards this goal, The Kavli Foundation convened a workshop in which participants began to outline how recent network science techniques can contextualize the emerging wave of neuroscientific big data, focusing on three topics: neurodevelopment, functional brain data, and health and disease. Below, we summarize these discussion points and outline opportunities by which the fields of network science and neuroscience can define common goals and language.

**Techniques**

Neuroscience

In the recent past, technical and experimental advancements in neuroscience have enabled scientists to study the brain at increasingly finer scales, ranging from coarse circuit analysis to whole-animal, cellular-level neural recording, connectivity mapping, and genetic profiling. While previous techniques already necessitated the use of graph theoretical tools, recent data collection methods have started to offer a consistent stream of multi-modal and high quality connectomic reconstructions that make the use of network science a necessity. For example, while a connectome of *C. elegans* has been available since 1986 (White et al. 1986), recent advances in electron microscopy (Abbott et al. 2020) have produced whole-animal wiring information in *Ciona Intestinalis (Ryan, Lu, and Meinertzhagen 2016)* and *Platynereis dumerilii (Verasztó et al. 2020)*, as well as brain-wide connectivity maps for *Drosophila* at different stages of development (Eichler et al. 2017; Scheffer et al. 2020; Winding et al. 2023), along with partial connectomes for zebrafish (Hildebrand et al. 2017), mice (Microns Consortium et al. 2021), and humans (Shapson-Coe et al. 2021). Single-cell transcriptomics has also enabled rapid and diverse profiling of cellular identity in various animals, developmental stages, and brain health patterns (Zeng 2022). Additional "bridge" techniques allow for rapid acquisition of multi-modal datasets, such as spatial transcriptomics (Chen et al. 2019), promising physical and genetic information of cells from a single measurement. These datasets offer detailed connectivity and identity information about thousands, and soon millions, of neurons. Analyzing this data, and extracting experimentally testable signals and hypotheses will be facilitated by integrating all data points via the use of network science tools, which in turn will also necessitate further advancement of current tools in network science.

To study functional properties of individual neurons and neural networks in the living brain, in vivo techniques such as two-photon microscopy (Grienberger and Konnerth 2012) and multi-unit electrode recording (Steinmetz et al. 2018) provide rapid profiling of local and mesoscale neuronal activity and anatomy in animal models, revealing principles of circuit organization and dynamic coding underlying a variety of neural processes in sensory perception, movement control, decision-making and behavior generation. However, presently the application of these approaches is largely restricted to one or a few brain regions at a time. Technology advancement is needed to monitor neuronal activities across multiple brain regions and at high resolution, necessary to truly understand the dynamic interplay of the different components of the brain-wide circuits for brain function.

Currently, magnetic resonance imaging (MRI) is the primary technology for noninvasively recording functioning brain networks in the human brain, either by reconstructing white matter tracts using diffusion tensor imaging (DTI) or by inferring axonal connectivity through the measurement of cytoarchitectural or morphometric similarity between brain regions. The rapid growth of datasets such as the Allen Human Brain Atlas (Hawrylycz et al. 2012) highlights the need for a wider range of human brain atlases that document gene expression and other molecular or cellular phenotypes that are commensurate with the structural phenotypes, such as volume and myelination. We must integrate multiple levels of analysis and apply diverse approaches, including computational and mathematical modeling, to successfully unravel the complexity of the brain networks and its many interacting components. Future functional brain profiling methods must also account for the multiple cell identities and network features that define neuronal systems.

> **Box 1: Common concepts in Network Neuroscience.**
> Studying the brain requires moving across scales and modalities, and a set of shared terms provides a common reference frame. **Structural Connectivity** refers to "ground-truth" physically instantiated networks, such as measured synaptic connections or axon tracts, while **Functional Connectivity** represents estimates of statistical dependencies between neural time courses. Similarly, **Dynamics *of* Networks** corresponds to changing of the connection topology, such as synaptic updates or pruning, while **Dynamics *on* Networks** refers to the neural activity patterns instantiated on top of the structural connectivity.

Network Science

Traditionally, neural connectivity is modeled as a simple graph, formalizing the brain as a set of nodes (neurons) connected by links (synapses, gap junctions). Network science is particularly well-suited to the study of such simple graphs: starting from the adjacency matrix of the system, encoding who is connected to whom, network science offers a suite of tools to characterize local and large scale structure, ranging from degree distributions to community structure, degree correlations, and even controllability, exploring our ability to guide the dynamics of the circuit. Yet, this "time-frozen" graph-based approach highly oversimplifies the true complexity of the brain, ignoring cell identities, signaling types, dynamics, and spatial and energetic constraints that shape this complex organ. Emerging approaches in network science offer a suite of tools to start capturing this rich complexity, helping us analyze the structural and functional brain data across scales.

For example, multiplex and multilayer networks provide a framework for understanding and describing cell-cell relationships and hierarchies, capturing the circuit motifs that can significantly impact the dynamical and topological properties of functional networks. Indeed, multiplex networks can represent multiple types of connections, such as synapses, gap junctions, neuromodulators, and circulating gut peptides, within a formal framework (Presigny and De Vico Fallani 2022; Bianconi 2018). Triadic interactions, in which a node affects the interaction between two other nodes, can also be incorporated, capturing for example how glia can influence the synaptic signal between neurons (Sun et al. 2023). These triadic interactions can lead to the emergence of higher-order networks, often represented as hypergraphs or simplicial complexes (Bianconi 2021; Battiston et al. 2020, 2021; Torres et al. 2021). Promise Theory furthers network analysis by incorporating complex agent modeling and conditional linkage, process interconnection language, and accounting for the functional and structural diversity of cells and their roles (Burgess 2021, 2015).

Finally, traditional network-based analyses of the brain have largely ignored the geometry and morphology of neurons, treating them as point-like nodes rather than physical objects with length, volume, and a branching tree structure. At larger scales, numerous network models have attempted to incorporate the physical dimension or geometry of extended neural networks, through considerations of wiring economy (Horvát et al. 2016; Markov et al. 2013), metabolic cost, and conduction delays (Bullmore and Sporns 2012). The emerging study of physical networks promises the tools to explore how the physicality and the spatial organization of the individual neurons and the non-crossing constraints affect the network structure of the whole brain (Pósfai et al. 2022). These approaches have the potential to address the metabolic cost of building and maintaining wiring, and incorporate the physical length of connections. There is a real need, for both network science and neuroscience, to go beyond simple connectivity information and incorporate the true physical nature of neurons, informed by weighing cell properties with their connections, allowing us to enrich our understanding of neuronal circuit operations.

**Application Areas**

<u>Neurodevelopment</u>

System neuroscience and genomics have relied on a fruitful collaboration between theory and experiment. However, a similar symbiosis has so far escaped neurodevelopment. Neurodevelopment has strong core principles, ripe for modeling, empowered by the recent availability of rich connectomics, genomics and imaging datasets, from which computational and network-based analyses can unleash rich insights. For instance, in addition to whole-body behavior and neural recordings, *C. elegans* now has developmentally resolved connectomes and transcriptomes (Boeck et al. 2016; Witvliet et al. 2021), allowing for the integrated analysis of connectivity, genetics, activity and behavior, inspiring the ongoing acquisition of similar datasets for larger organisms. The main use of network tools in brain science has so far been limited to the mapping and analysis of static network maps. However, a key discovery of network science is that we must understand the regularities and rules governing the growth and assembly of networks, i.e., the evolving topology of their connectivity, to understand the origin of the empirically observed network characteristics (Barabasi and Albert 1999). Network science offers important tools to address this gap, and hence can provide a comprehensive quantitative framework to study and understand neurodevelopment. It offers the formal language to describe, and then to analyze, how the emerging cell identity and its physical instantiation leads to the observed connectivity relationships in the brain and ultimately shapes their impact on brain function. Reciprocally, new insights from biological systems that first establish and then prune structured networks may inspire new network approaches (Woźniak et al. 2020).

One central question in this field is how neuron identity, captured by gene expression profiles, location, and shape, determines the wiring patterns of neurons and leads to stereotyped connectivity and behavior. Network models of neurodevelopmental principles are needed, therefore, to validate hypotheses and make predictions for future experiments. For instance, Roger Sperry's hypothesis that genetic compatibility drives neuronal connectivity, helped infer the protein interactions that underlie connectivity in *C. elegans (Barabási and Barabási 2020; Kovács, Barabási, and Barabási 2020)* and in the *Drosophila* visual system (Kurmangaliyev et al. 2020). These models are most successful when they take into account the affordances of the niche in which organisms operate, including noise from data collection limitations and spatial restrictions, offering more accurate descriptions of the complex landscape of neuronal circuit construction.

Further work on cell migration, morphogenesis and axon guidance can help unveil the developmental constraints that lead to specific circuit implementations and overall network assembly. For example, the preconfigured dynamics of the hippocampus have been shown to be influenced by factors such as embryonic birthdate and neurogenesis rate (Huszár et al. 2022). Additionally, it is now known that certain network features, including heavy-tailed degree distributions, modularity, and interconnected hubs, are present across species and scales in the brain (van den Heuvel, Bullmore, and Sporns 2016; Buzsáki and Mizuseki 2014; Towlson et al. 2013). One potential explanation for the conservation of these features is the existence of universal constraints on the brain's physical architecture that arise from the trade-off between the cost of development, physical constraints and coding efficiency. In this context, it is likely that high-cost components, such as long-distance inter-modular tracts, are topologically integrative in order to minimize the transmission time of signals between spatially distant brain regions (Bullmore and Sporns 2012). Further research in this area has the potential to improve our understanding of the development and organization of the brain, with potential implications for the diagnosis and treatment of neurological disorders, as we discuss later.

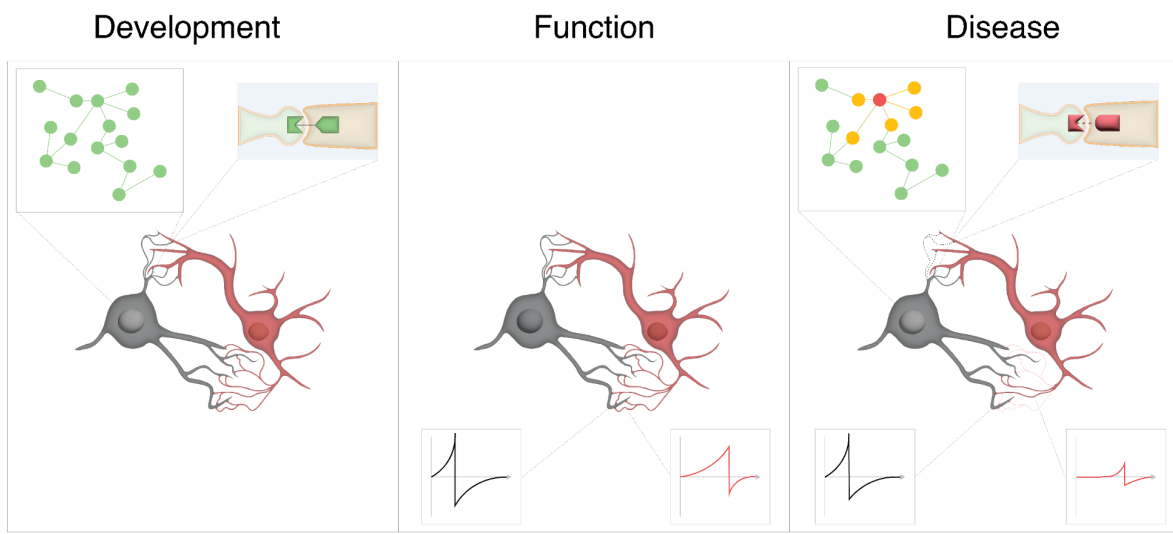

**Figure 1: Multi-scale interaction in network development, function and disease.**
**Development:** Neural connectivity emerges as a function of cell identity, linking network dynamics across modalities and scales. Regulatory networks (top left) underlie cell differentiation, and protein-protein interactions guide morphological maturation and synaptic specificity (top right).
**Function:** Structural connectivity guides the emergent possibilities of functional networks, determining the strength with which one neuron can influence the next (bottom).
**Disease:** In a diseased state, failures at multiple network levels leads to perturbed function. Genetic mutations cause disruptions in gene regulatory networks (top left), as well as conformation changes that change protein-protein interactions (top right), potentially leading to loss of synaptic connectivity (dashed neurites). In turn, reduced connection strength between neurons disruptions activity propagation (bottom), providing links between genetic changes and cognitive dysfunction.

Brains are Networks that Do

In technological networks such as the Internet or a computer chip, structure and function are carefully separated: information is encoded into the signal, hence the role of the network is only to guarantee routing paths between nodes. In the brain, however, action potentials do not encode information in isolation. Instead, the brain relies on population coding, meaning that encoding is implemented by the patterns of signals generated by multiple *physical networks* of connections. Thus, monitoring and quantifying this *network structure* is critical for understanding how neuronal coding achieves information processing. This makes the structure of the network more than a propagation backbone; it becomes an integral part of the algorithm itself (Molnár et al. 2020). Thus, the connectome cannot be understood divorced from the context of the actions it performs. Hence, the modules, metrics, and generative processes that support robust representation are needed to be integrated with the structural representation.

While many recent studies have revealed ways in which task structures are reflected in the networks of neural representations (Chung and Abbott 2021), little work has been done to elucidate how such representational geometries arise mechanistically and dynamically. Future research should aim to unveil how connectivity patterns at multiple scales influence the population-level representational geometry, and how this leads to the implementation of behaviorally relevant task structures. For instance, hippocampal "cognitive maps" that support reasoning in different encountered task spaces have a natural extension to network formalism: each behavioral state can be a node, and possible transitions between states are edges (Muller 1996; Eichenbaum and Cohen 2004; Stachenfeld, Botvinick, and Gershman 2017; George et al. 2021), a representation that can be extended to the challenge of inducing latent networks from sequential inputs (Raju et al. 2022). Further, we must account for the dual dynamics present in the brain: network connectivity defines the possible functions that can be supported. In the reverse direction, the functional dynamics of the network allows synapses to form and change, allowing dynamics (activity) to change the connectivity of the underlying networks (Papadimitriou et al. 2020). Important insights into brain function can be revealed when dynamics taking place at the node level (single neurons, brain regions) are integrated with dynamics taking place on links (synaptic signals, edge signals) (Faskowitz, Betzel, and Sporns

2022) or on higher-order motifs (Santoro et al. 2023) which are driven by the network cyclic structure and its higher-order topology (Millán, Torres, and Bianconi 2020). For instance, studying symmetrical complexes, such as automorphisms (Morone and Makse 2019) and fibrations (Morone, Leifer, and Makse 2020), in structural and functional neural connectivities has succeeded in unveiling the building blocks for neural synchronization in the brain. Graph neural networks (Battaglia et al. 2018; Bronstein et al. 2021), which combine the benefits of network topology and machine learning, may also help us relate connectomically constrained graphs to the neural dynamics that take place over them.

For a brain to carry out the numerous processes it supports, it is expected to simultaneously control the activity of the individual neurons, as well as the dynamics of individual circuits and ultimately the full network. This represents an enormously complex control task, as unveiled by recent advances in network controllability that merged the tools of control theory and network science (Liu and Barabási 2016). These tools help us identify the nodes through which one can control a complex neural circuit, just like a car is controlled through three core mechanisms: the steering wheel, gas pedal and brake. Recent work used network control to predict the function of individual neurons in the *C. elegans* connectome, leading to the discovery of new neurons involved in the control of locomotion, and offering direct falsifiable experimental confirmation of control principles (Yan et al. 2017). An alternate description of brain function requires a deeper understanding of the underlying control problems, which requires simultaneous profiling and understanding of network structure and dynamics (Stiso et al. 2019; Tang and Bassett 2018).

Finally, machine learning methods have offered a unique approach for linking network structure to task performance (Veličković 2023; Chami et al. 2020), allowing for rapid profiling of learning and behavior that can later inform how we query biological learning (Marblestone, Wayne, and Kording 2016; Vu et al. 2018; Richards et al. 2019). To move forward, we must study the statistics of AI architecture's weight structures that offer high performance on complex tasks, helping identify powerful subnetworks, or "winning tickets," responsible for the majority of the performance of a system. An alternative approach lies in identifying generative processes that produce highly performing networks. This is inspired by innate behaviors: animals arrive into the world with a set of evolution-tested preexisting dynamics, implying an optimized set of developmental processes that yield a fine-tuned functional connectome at birth (Zador 2019; Barabási, Schuhknecht, and Engert 2022). This process, termed the "genomic bottleneck," has the potential to greatly increase the flexibility and utility of AI systems (Koulakov et al. 2022). Indeed, developmentally inspired encodings of neural network weights have already shown high and stable performance on reinforcement learning, metalearning and transfer learning tasks (Barabási, Beynon, and Katona 2022).

Further work in these directions would require streamlined integrations of powerful circuits identified in the connectome with machine learning systems. A major barrier lies in the complexity of the initial setup of tasks that the networks are asked to learn (Seshadhri et al. 2020), embedded in complex packages like the simulated physics environment of Mujoco (Todorov, Erez, and Tassa 2012). It is also challenging to provide custom topologies or weights to current machine learning packages, thereby moving past the standard feed-forward, layered architectures. Addressing these challenges would allow the network science toolkit to define a systematic search of network priors in machine learning, thereby modeling the neuroevolutionary processes and neurodevelopmental solutions responsible for biological intelligence.

Health and Disease

The integration of multiple data streams is crucial for understanding the neural transitions from a healthy state to a disease state, particularly in the context of brain disorders, diseases, and mental illnesses, often rooted in the early years of life. Large-scale MRI datasets have allowed for the modeling of normative trajectories of brain development (Bethlehem et al. 2022), however, major opportunities remain for network science to reveal the causes and physiologies of brain disorders through population analyses.

In addition to the brain's own internal networks, the connections between the brain and other organs robustly affect neural development and function. Complex interactions have been revealed in the gut-brain axis, where microbiota can modulate immune and neural states, as well as in the brain's interaction with the reproductive system, driving intricate fluctuations in levels of sex hormones during puberty, menopause, and pregnancy (Pritschet et al. 2020; Andreano et al. 2018). Overall, the connections between the brain and other organs can have significant effects on neural development and function, highlighting the importance of a holistic exploration of neural networks together with the body as a whole (Buzsáki and Tingley 2023).

Ultimately, to diagnose and treat disease, we must understand the complex interactions between genetic, disease, and drug networks and their impact on the connectome. Toward that goal, Network Neuroscience must partner with Network Medicine, which applies network science to subcellular interactions, aiming to diagnose, prevent, and treat diseases. This need is reinforced by studies that have found that high degree hubs, located mainly in dorsolateral prefrontal, lateral and medial temporal, and cingulate areas of human cortex, are co-located with an enrichment of neurodevelopmental and neurotransmitter-related genes and implicated in the pathogenesis of schizophrenia (Morgan et al. 2019). Network Medicine takes advantage of the structure of subcellular networks, as captured by experimentally mapped protein and noncoding interactions, to identify disease mechanisms, therapeutic targets, drug-repurposing opportunities, and biomarkers. In the case of brain diseases, mutations and other molecular changes that alter the subcellular networks within neurons and non-neuronal cells, in turn affect the wiring and rewiring of the connectome and neural dynamics. Hence, effective interventions and treatments for brain disorders must confront the double network problem, accounting for the impact of changes in the subcellular network on connectivity and ultimately brain function.

**Conclusions**

Major funding directives, like the public-private funding alliance of the US BRAIN Initiative, have significantly advanced the development of technologies for studying the brain across scales and modalities. Yet, the massive amount of data produced and expected to emerge from these tools have created a complexity bottleneck. We need guiding frameworks to organize and conceptualize these data, leading to falsifiable hypotheses. Network science offers a natural match for this task, with the potential for integrating complexity across cell identities, signaling types, dynamics, and spatial and energetic constraints that shape brain development, function and disease.

We need the joint engagement of network scientists and neuroscientists, to develop novel network-based methods tailored to brain science, in order to address the unique priorities and challenges posed by brain research. Network-based methods will need to account for the dynamic nature of connections in the brain, which are continually changing as a result of various factors such as experience, aging, and disease, as well as incomplete or uncertain reconstructions of brain connectivity. Continued advances in neuroscience have opened up exciting possibilities for a deeper understanding of the brain and its function, and now require network science to capture the dynamics of this complex system with the goals of unlocking how neural identity, dynamics, behavior and disease all link together.

These methodological advances can run parallel to ever-increasing efforts towards adoption of open-science practices such as data and code sharing. Such efforts bring new challenges related to reproducibility, and have, in some cases, resulted in examples of findings that fail to replicate (Open Science Collaboration 2015; Errington et al. 2021) or exhibit substantial variability attributable to software (Botvinik-Nezer et al. 2020; Bowring, Maumet, and Nichols 2019) or analysis teams (Botvinik-Nezer et al. 2020). As a discipline, neuroimaging has championed open science initiatives, promoting practices including detailed methodological descriptions and sharing of data and code used to generate results in a publication (Nichols et al. 2016), and even multiverse analyses that consider all plausible analytical variations (Dafflon et al. 2022).

To achieve these goals, there is a need to facilitate greater interaction between the network science and neuroscience communities. A well-tested way is to offer interdisciplinary grants from public and private organizations, such as The Kavli Foundation, the NIH and the NSF, that focus on developing network tools for emerging neuroscience technologies and questions, as well as support for students and postdoctoral fellows with interests in both disciplines. These grants could also support workshops and conferences that bring together researchers from both fields, and provide funding for coursework in network neuroscience at the undergraduate and graduate levels. Actively fostering collaboration between these two fields will encourage the development of integrative approaches to understanding biological data, a necessary step towards advancing our understanding of the brain and its functions.

**Acknowledgements**

The authors would like to thank The Kavli Foundation for organizing and supporting a convening in October 2022, "Network Science Meets Neuroscience," which inspired this review, and Daria Koshkina for designing figures. G.B. acknowledges support from the Turing-Roche partnership and the Royal Society (IEC\NSFC\191147). H.M. acknowledge support from NIBIB and NIMH through the NIH BRAIN Initiative Grant # R01 EB028157. E.K.T. acknowledges funding from the Government of Canada's New Frontiers in Research Fund (NFRF), NFRFE-2021-00420, and the Natural Sciences and Engineering Research Council of Canada (NSERC), funding reference number RGPIN-2021-02949. H.Z. acknowledges support from NIH BRAIN Initiative grant U19MH114830.

**Author Contributions**
D.L.B., A.-L.B., and Gy.B. wrote the first draft. D.L.B., A.-L.B., A.B., and Gy.B. coordinated authors and provided extensive editing to successive drafts. All authors provided additional guidance and edits at various stages.

**References**

Abbott, Larry F., Davi D. Bock, Edward M. Callaway, Winfried Denk, Catherine Dulac, Adrienne L. Fairhall, Ila Fiete, et al. 2020. "The Mind of a Mouse." *Cell* 182 (6): 1372–76.

Andreano, Joseph M., Alexandra Touroutoglou, Brad Dickerson, and Lisa Feldman Barrett. 2018. "Hormonal Cycles, Brain Network Connectivity, and Windows of Vulnerability to Affective Disorder." *Trends in Neurosciences* 41 (10): 660–76.

Barabasi, A. L., and R. Albert. 1999. "Emergence of Scaling in Random Networks." *Science* 286 (5439): 509–12.

Barabási, Dániel L., and Albert-László Barabási. 2020. "A Genetic Model of the Connectome." *Neuron* 105 (3): 435–45.e5.

Barabási, Dániel L., Taliesin Beynon, and Ádám Katona. 2022. "Complex Computation from Developmental Priors." *Biorxiv*. https://doi.org/10.1101/2021.03.29.437584.

Barabási, Dániel L., Gregor F. P. Schuhknecht, and Florian Engert. 2022. "Nature over Nurture: Functional Neuronal Circuits Emerge in the Absence of Developmental Activity." *Biorxiv*. https://doi.org/10.1101/2022.10.24.513526.

Bassett, Danielle S., Perry Zurn, and Joshua I. Gold. 2018. "On the Nature and Use of Models in Network Neuroscience." *Nature Reviews. Neuroscience* 19 (9): 566–78.

Battaglia, Peter W., Jessica B. Hamrick, Victor Bapst, Alvaro Sanchez-Gonzalez, Vinicius Zambaldi, Mateusz Malinowski, Andrea Tacchetti, et al. 2018. "Relational Inductive Biases, Deep

Learning, and Graph Networks." https://doi.org/10.48550/ARXIV.1806.01261.

Battiston, Federico, Enrico Amico, Alain Barrat, Ginestra Bianconi, Guilherme Ferraz de Arruda, Benedetta Franceschiello, Iacopo Iacopini, et al. 2021. "The Physics of Higher-Order Interactions in Complex Systems." *Nature Physics*. https://doi.org/10.1038/s41567-021-01371-4.

Battiston, Federico, Giulia Cencetti, Iacopo Iacopini, Vito Latora, Maxime Lucas, Alice Patania, Jean-Gabriel Young, and Giovanni Petri. 2020. "Networks beyond Pairwise Interactions: Structure and Dynamics." *Physics Reports*. https://doi.org/10.1016/j.physrep.2020.05.004.

Bethlehem, R. A. I., J. Seidlitz, S. R. White, J. W. Vogel, K. M. Anderson, C. Adamson, S. Adler, et al. 2022. "Brain Charts for the Human Lifespan." *Nature* 604 (7906): 525–33.

Bianconi, Ginestra. 2018. "Multilayer Networks." *Oxford Scholarship Online*. https://doi.org/10.1093/oso/9780198753919.001.0001.

———. 2021. "Higher-Order Networks." https://doi.org/10.1017/9781108770996.

Boeck, Max E., Chau Huynh, Lou Gevirtzman, Owen A. Thompson, Guilin Wang, Dionna M. Kasper, Valerie Reinke, Ladeana W. Hillier, and Robert H. Waterston. 2016. "The Time-Resolved Transcriptome of C. Elegans." *Genome Research* 26 (10): 1441–50.

Botvinik-Nezer, Rotem, Felix Holzmeister, Colin F. Camerer, Anna Dreber, Juergen Huber, Magnus Johannesson, Michael Kirchler, et al. 2020. "Variability in the Analysis of a Single Neuroimaging Dataset by Many Teams." *Nature* 582 (7810): 84–88.

Bowring, Alexander, Camille Maumet, and Thomas E. Nichols. 2019. "Exploring the Impact of Analysis Software on Task fMRI Results." *Human Brain Mapping* 40 (11): 3362–84.

Bronstein, Michael M., Joan Bruna, Taco Cohen, and Petar Veličković. 2021. "Geometric Deep Learning: Grids, Groups, Graphs, Geodesics, and Gauges." https://doi.org/10.48550/ARXIV.2104.13478.

Bullmore, Ed, and Olaf Sporns. 2012. "The Economy of Brain Network Organization." *Nature Reviews Neuroscience*. https://doi.org/10.1038/nrn3214.

Burgess, Mark. 2015. "Spacetimes with Semantics (II), Scaling of Agency, Semantics, and Tenancy." https://doi.org/10.48550/ARXIV.1505.01716.

———. 2021. "Motion of the Third Kind (I) Notes on the Causal Structure of Virtual Processes for Privileged Observers." Unpublished. https://doi.org/10.13140/RG.2.2.30483.35361.

Buzsáki, György, and Kenji Mizuseki. 2014. "The Log-Dynamic Brain: How Skewed Distributions Affect Network Operations." *Nature Reviews. Neuroscience* 15 (4): 264–78.

Buzsáki, György, and David Tingley. 2023. "Cognition from the Body-Brain Partnership: Exaptation of Memory." *Annual Review of Neuroscience*, March. https://doi.org/10.1146/annurev-neuro-101222-110632.

Chami, Ines, Sami Abu-El-Haija, Bryan Perozzi, Christopher Ré, and Kevin Murphy. 2020. "Machine Learning on Graphs: A Model and Comprehensive Taxonomy," May. https://doi.org/10.48550/arXiv.2005.03675.

Chen, Xiaoyin, Yu-Chi Sun, Huiqing Zhan, Justus M. Kebschull, Stephan Fischer, Katherine Matho, Z. Josh Huang, Jesse Gillis, and Anthony M. Zador. 2019. "High-Throughput Mapping of Long-Range Neuronal Projection Using In Situ Sequencing." *Cell* 179 (3): 772–86.e19.

Chung, Sueyeon, and L. F. Abbott. 2021. "Neural Population Geometry: An Approach for Understanding Biological and Artificial Neural Networks." *Current Opinion in Neurobiology* 70 (October): 137–44.

Dafflon, Jessica, Pedro F Da Costa, František Váša, Ricardo Pio Monti, Danilo Bzdok, Peter J. Hellyer, Federico Turkheimer, Jonathan Smallwood, Emily Jones, and Robert Leech. 2022. "A Guided Multiverse Study of Neuroimaging Analyses." *Nature Communications* 13 (1): 3758.

Eichenbaum, Howard, and Neal J. Cohen. 2004. *From Conditioning to Conscious Recollection: Memory Systems of the Brain*. Oxford University Press on Demand.

Eichler, Katharina, Feng Li, Ashok Litwin-Kumar, Youngser Park, Ingrid Andrade, Casey M. Schneider-Mizell, Timo Saumweber, et al. 2017. "The Complete Connectome of a Learning and Memory Centre in an Insect Brain." *Nature* 548 (7666): 175–82.

Errington, Timothy M., Alexandria Denis, Nicole Perfito, Elizabeth Iorns, and Brian A. Nosek. 2021. "Challenges for Assessing Replicability in Preclinical Cancer Biology." *eLife* 10 (December). https://doi.org/10.7554/eLife.67995.

Faskowitz, Joshua, Richard F. Betzel, and Olaf Sporns. 2022. "Edges in Brain Networks: Contributions to Models of Structure and Function." *Network Neuroscience (Cambridge, Mass.)* 6 (1): 1–28.

Fornito, Alex, Andrew Zalesky, and Michael Breakspear. 2015. "The Connectomics of Brain Disorders." *Nature Reviews. Neuroscience* 16 (3): 159–72.

George, Dileep, Rajeev V. Rikhye, Nishad Gothoskar, J. Swaroop Guntupalli, Antoine Dedieu, and Miguel Lázaro-Gredilla. 2021. "Clone-Structured Graph Representations Enable Flexible Learning and Vicarious Evaluation of Cognitive Maps." *Nature Communications* 12 (1): 2392.

Grienberger, Christine, and Arthur Konnerth. 2012. "Imaging Calcium in Neurons." *Neuron* 73 (5): 862–85.

Hawrylycz, Michael J., Ed S. Lein, Angela L. Guillozet-Bongaarts, Elaine H. Shen, Lydia Ng, Jeremy A. Miller, Louie N. van de Lagemaat, et al. 2012. "An Anatomically Comprehensive Atlas of the Adult Human Brain Transcriptome." *Nature* 489 (7416): 391–99.

Heuvel, Martijn P. van den, Edward T. Bullmore, and Olaf Sporns. 2016. "Comparative Connectomics." *Trends in Cognitive Sciences*. https://doi.org/10.1016/j.tics.2016.03.001.

Hildebrand, David Grant Colburn, Marcelo Cicconet, Russel Miguel Torres, Woohyuk Choi, Tran Minh Quan, Jungmin Moon, Arthur Willis Wetzel, et al. 2017. "Whole-Brain Serial-Section Electron Microscopy in Larval Zebrafish." *Nature* 545 (7654): 345–49.

Horvát, Szabolcs, Răzvan Gămănuț, Mária Ercsey-Ravasz, Loïc Magrou, Bianca Gămănuț, David C. Van Essen, Andreas Burkhalter, Kenneth Knoblauch, Zoltán Toroczkai, and Henry Kennedy. 2016. "Spatial Embedding and Wiring Cost Constrain the Functional Layout of the Cortical Network of Rodents and Primates." *PLoS Biology* 14 (7): e1002512.

Huszár, Roman, Yunchang Zhang, Heike Blockus, and György Buzsáki. 2022. "Preconfigured Dynamics in the Hippocampus Are Guided by Embryonic Birthdate and Rate of Neurogenesis." *Nature Neuroscience* 25 (9): 1201–12.

Koulakov, Alexei, Sergey Shuvaev, Divyansha Lachi, and Anthony Zador. 2022. "Encoding Innate Ability through a Genomic Bottleneck." *Biorxiv*. https://doi.org/10.1101/2021.03.16.435261.

Kovács, István A., Dániel L. Barabási, and Albert-László Barabási. 2020. "Uncovering the Genetic Blueprint of the Nervous System." *Proceedings of the National Academy of Sciences of the United States of America* 117 (52): 33570–77.

Kurmangaliyev, Yerbol Z., Juyoun Yoo, Javier Valdes-Aleman, Piero Sanfilippo, and S. Lawrence Zipursky. 2020. "Transcriptional Programs of Circuit Assembly in the Drosophila Visual System." *Neuron* 108 (6): 1045–57.e6.

Liu, Yang-Yu, and Albert-László Barabási. 2016. "Control Principles of Complex Systems." *Reviews of Modern Physics* 88 (3). https://doi.org/10.1103/revmodphys.88.035006.

Marblestone, Adam H., Greg Wayne, and Konrad P. Kording. 2016. "Toward an Integration of Deep Learning and Neuroscience." *Frontiers in Computational Neuroscience* 10 (September): 94.

Markov, Nikola T., Mária Ercsey-Ravasz, David C. Van Essen, Kenneth Knoblauch, Zoltán Toroczkai, and Henry Kennedy. 2013. "Cortical High-Density Counterstream Architectures." *Science* 342 (6158): 1238406.

Microns Consortium, MICrONS Consortium, J. Alexander Bae, Mahaly Baptiste, Agnes L. Bodor, Derrick Brittain, Joann Buchanan, et al. 2021. "Functional Connectomics Spanning Multiple Areas of Mouse Visual Cortex." *Biorxiv*. https://doi.org/10.1101/2021.07.28.454025.

Millán, Ana P., Joaquín J. Torres, and Ginestra Bianconi. 2020. "Explosive Higher-Order Kuramoto Dynamics on Simplicial Complexes." *Physical Review Letters* 124 (21): 218301.

Molnár, Ferenc, Shubha R. Kharel, Xiaobo Sharon Hu, and Zoltán Toroczkai. 2020. "Accelerating a Continuous-Time Analog SAT Solver Using GPUs." *Computer Physics Communications* 256 (107469): 107469.

Morgan, Sarah E., Jakob Seidlitz, Kirstie J. Whitaker, Rafael Romero-Garcia, Nicholas E. Clifton, Cristina Scarpazza, Therese van Amelsvoort, et al. 2019. "Cortical Patterning of Abnormal Morphometric Similarity in Psychosis Is Associated with Brain Expression of Schizophrenia-Related Genes." *Proceedings of the National Academy of Sciences*. https://doi.org/10.1073/pnas.1820754116.

Morone, Flaviano, Ian Leifer, and Hernán A. Makse. 2020. "Fibration Symmetries Uncover the Building Blocks of Biological Networks." *Proceedings of the National Academy of Sciences of the United States of America* 117 (15): 8306–14.

Morone, Flaviano, and Hernán A. Makse. 2019. "Symmetry Group Factorization Reveals the Structure-Function Relation in the Neural Connectome of Caenorhabditis Elegans." *Nature Communications* 10 (1): 4961.

Muller, R. 1996. "A Quarter of a Century of Place Cells." *Neuron* 17 (5): 813–22.

Nichols, Thomas E., Samir Das, Simon B. Eickhoff, Alan C. Evans, Tristan Glatard, Michael Hanke, Nikolaus Kriegeskorte, et al. 2016. "Best Practices in Data Analysis and Sharing in Neuroimaging Using MRI." *bioRxiv*. bioRxiv. https://doi.org/10.1101/054262.

Open Science Collaboration. 2015. "PSYCHOLOGY. Estimating the Reproducibility of Psychological Science." *Science* 349 (6251): aac4716.

Papadimitriou, Christos H., Santosh S. Vempala, Daniel Mitropolsky, Michael Collins, and Wolfgang Maass. 2020. "Brain Computation by Assemblies of Neurons." *Proceedings of the National Academy of Sciences of the United States of America* 117 (25): 14464–72.

Pósfai, Márton, Balázs Szegedy, Iva Bačić, Luka Blagojević, Miklós Abért, János Kertész, László Lovász, and Albert-László Barabási. 2022. "Understanding the Impact of Physicality on Network Structure." https://doi.org/10.48550/ARXIV.2211.13265.

Presigny, Charley, and Fabrizio De Vico Fallani. 2022. "*Colloquium* : Multiscale Modeling of Brain Network Organization." *Reviews of Modern Physics*.

https://doi.org/10.1103/revmodphys.94.031002.

Pritschet, Laura, Tyler Santander, Caitlin M. Taylor, Evan Layher, Shuying Yu, Michael B. Miller, Scott T. Grafton, and Emily G. Jacobs. 2020. "Functional Reorganization of Brain Networks across the Human Menstrual Cycle." *NeuroImage* 220 (October): 117091.

Pulvermüller, Friedemann, Rosario Tomasello, Malte R. Henningsen-Schomers, and Thomas Wennekers. 2021. "Biological Constraints on Neural Network Models of Cognitive Function." *Nature Reviews. Neuroscience* 22 (8): 488–502.

Raju, Rajkumar Vasudeva, J. Swaroop Guntupalli, Guangyao Zhou, Miguel Lázaro-Gredilla, and Dileep George. 2022. "Space Is a Latent Sequence: Structured Sequence Learning as a Unified Theory of Representation in the Hippocampus." https://doi.org/10.48550/ARXIV.2212.01508.

Richards, Blake A., Timothy P. Lillicrap, Philippe Beaudoin, Yoshua Bengio, Rafal Bogacz, Amelia Christensen, Claudia Clopath, et al. 2019. "A Deep Learning Framework for Neuroscience." *Nature Neuroscience* 22 (11): 1761–70.

Ryan, Kerrianne, Zhiyuan Lu, and Ian A. Meinertzhagen. 2016. "The CNS Connectome of a Tadpole Larva of (L.) Highlights Sidedness in the Brain of a Chordate Sibling." *eLife* 5 (December). https://doi.org/10.7554/eLife.16962.

Santoro, Andrea, Federico Battiston, Giovanni Petri, and Enrico Amico. 2023. "Higher-Order Organization of Multivariate Time Series." *Nature Physics*, January. https://doi.org/10.1038/s41567-022-01852-0.

Scheffer, Louis K., C. Shan Xu, Michal Januszewski, Zhiyuan Lu, Shin-Ya Takemura, Kenneth J. Hayworth, Gary B. Huang, et al. 2020. "A Connectome and Analysis of the Adult Central Brain." *eLife* 9 (September). https://doi.org/10.7554/eLife.57443.

Seshadhri, C., Aneesh Sharma, Andrew Stolman, and Ashish Goel. 2020. "The Impossibility of Low-Rank Representations for Triangle-Rich Complex Networks." *Proceedings of the National Academy of Sciences of the United States of America* 117 (11): 5631–37.

Shapson-Coe, Alexander, Michał Januszewski, Daniel R. Berger, Art Pope, Yuelong Wu, Tim Blakely, Richard L. Schalek, et al. 2021. "A Connectomic Study of a Petascale Fragment of Human Cerebral Cortex." *Biorxiv*. https://doi.org/10.1101/2021.05.29.446289.

Stachenfeld, Kimberly L., Matthew M. Botvinick, and Samuel J. Gershman. 2017. "The Hippocampus as a Predictive Map." *Nature Neuroscience*. https://doi.org/10.1038/nn.4650.

Steinmetz, Nicholas A., Christof Koch, Kenneth D. Harris, and Matteo Carandini. 2018. "Challenges and Opportunities for Large-Scale Electrophysiology with Neuropixels Probes." *Current Opinion in Neurobiology* 50 (June): 92–100.

Stiso, Jennifer, Ankit N. Khambhati, Tommaso Menara, Ari E. Kahn, Joel M. Stein, Sandihitsu R. Das, Richard Gorniak, et al. 2019. "White Matter Network Architecture Guides Direct Electrical Stimulation through Optimal State Transitions." *Cell Reports* 28 (10): 2554–66.e7.

Sun, Hanlin, Filippo Radicchi, Jürgen Kurths, and Ginestra Bianconi. 2023. "The Dynamic Nature of Percolation on Networks with Triadic Interactions." *Nature Communications* 14 (1): 1308.

Tang, Evelyn, and Danielle S. Bassett. 2018. "Colloquium: Control of Dynamics in Brain Networks." *Reviews of Modern Physics* 90 (3). https://doi.org/10.1103/revmodphys.90.031003.

Todorov, Emanuel, Tom Erez, and Yuval Tassa. 2012. "MuJoCo: A Physics Engine for Model-Based Control." *2012 IEEE/RSJ International Conference on Intelligent Robots and Systems*. https://doi.org/10.1109/iros.2012.6386109.

Torres, Leo, Ann S. Blevins, Danielle Bassett, and Tina Eliassi-Rad. 2021. "The Why, How, and When of Representations for Complex Systems." *SIAM Review*. https://doi.org/10.1137/20m1355896.

Towlson, Emma K., Petra E. Vértes, Sebastian E. Ahnert, William R. Schafer, and Edward T. Bullmore. 2013. "The Rich Club of the C. Elegans Neuronal Connectome." *The Journal of Neuroscience: The Official Journal of the Society for Neuroscience* 33 (15): 6380–87.

Veličković, Petar. 2023. "Everything Is Connected: Graph Neural Networks." *Current Opinion in Structural Biology* 79 (February): 102538.

Verasztó, Csaba, Sanja Jasek, Martin Gühmann, Réza Shahidi, Nobuo Ueda, James David Beard, Sara Mendes, et al. 2020. "Whole-Animal Connectome and Cell-Type Complement of the Three-Segmented Platynereis Dumerilii Larva." *bioRxiv*. bioRxiv. https://doi.org/10.1101/2020.08.21.260984.

Vu, Mai-Anh T., Tülay Adalı, Demba Ba, György Buzsáki, David Carlson, Katherine Heller, Conor Liston, et al. 2018. "A Shared Vision for Machine Learning in Neuroscience." *The Journal of Neuroscience: The Official Journal of the Society for Neuroscience* 38 (7): 1601–7.

White, J. G., E. Southgate, J. N. Thomson, and S. Brenner. 1986. "The Structure of the Nervous System of the Nematode Caenorhabditis Elegans." *Philosophical Transactions of the Royal Society of London. Series B, Biological Sciences* 314 (1165): 1–340.

Winding, Michael, Benjamin D. Pedigo, Christopher L. Barnes, Heather G. Patsolic, Youngser Park, Tom Kazimiers, Akira Fushiki, et al. 2023. "The Connectome of an Insect Brain." *Science* 379 (6636): eadd9330.

Witvliet, Daniel, Ben Mulcahy, James K. Mitchell, Yaron Meirovitch, Daniel R. Berger, Yuelong Wu, Yufang Liu, et al. 2021. "Connectomes across Development Reveal Principles of Brain Maturation." *Nature* 596 (7871): 257–61.

Woźniak, Stanisław, Angeliki Pantazi, Thomas Bohnstingl, and Evangelos Eleftheriou. 2020. "Deep Learning Incorporating Biologically Inspired Neural Dynamics and in-Memory Computing." *Nature Machine Intelligence* 2 (6): 325–36.

Yan, Gang, Petra E. Vértes, Emma K. Towlson, Yee Lian Chew, Denise S. Walker, William R. Schafer, and Albert-László Barabási. 2017. "Network Control Principles Predict Neuron Function in the Caenorhabditis Elegans Connectome." *Nature* 550 (7677): 519–23.

Zador, Anthony M. 2019. "A Critique of Pure Learning and What Artificial Neural Networks Can Learn from Animal Brains." *Nature Communications* 10 (1): 3770.

Zeng, Hongkui. 2022. "What Is a Cell Type and How to Define It?" *Cell*. https://doi.org/10.1016/j.cell.2022.06.031.